\documentclass[reprint,superscriptaddress,aps,prb]{revtex4-2}

\usepackage{amsmath,epsfig,amssymb,subfigure,bm,dsfont}
\usepackage{lipsum} 

\usepackage{color}
\usepackage{graphicx}% Include figure files
\usepackage{dcolumn}% Align table columns on decimal point
\usepackage{bm}% bold math
\usepackage{bbold}% bbold package
\usepackage{upgreek}% to have up mu
\usepackage{amsfonts,color}
\usepackage{amsmath}
\usepackage{float}
\usepackage{natbib}
\setcitestyle{articletitle=true}

\usepackage[unicode=true, breaklinks=false, pdfborder={0 0 1}, backref=false, colorlinks=true, linkcolor=blue, urlcolor=blue, citecolor=blue]{hyperref}

\newcounter{lastnote}

\begin{document}

\title{Disambiguating electrical detection of magnetization dynamics in magnetic insulators}

\author{Hanchen Wang}
\thanks{These authors contributed equally to this work.}
\email{hanchen.wang@mat.ethz.ch}
\affiliation{%
Department of Materials, ETH Zurich, Zurich 8093, Switzerland
}%

\author{William~Legrand}
\thanks{These authors contributed equally to this work.}
\email{william.legrand@neel.cnrs.fr}
\affiliation{%
Universit\'e Grenoble Alpes, CNRS, Institut N\'eel, Grenoble 38042, France
}%

\author{Shangyuan~Wang}
\thanks{These authors contributed equally to this work.}
\affiliation{%
The Center for Advanced Quantum Studies and School of Physics and Astronomy, Beijing Normal University, Beijing, 100875, China.
}%
\affiliation{%
Key Laboratory of Multiscale Physics, Ministry of Education, Beijing Normal University, Beijing, 100875, China.
}%

\author{Davit Petrosyan}
\affiliation{%
Department of Materials, ETH Zurich, Zurich 8093, Switzerland
}%

\author{Hiroki~Matsumoto}
\affiliation{%
Department of Materials, ETH Zurich, Zurich 8093, Switzerland
}%
\affiliation{Institute for Chemical Research, Kyoto University, 6110011 Uji, Japan}

\author{Alexander~E.~Kossak}
\affiliation{%
Department of Materials, ETH Zurich, Zurich 8093, Switzerland
}%

\author{Richard~Schlitz}
\affiliation{Department of Physics, University of Konstanz, 78457 Konstanz, Germany}

\author{Ka~Shen}
\email{kashen@bnu.edu.cn}
\affiliation{%
The Center for Advanced Quantum Studies and School of Physics and Astronomy, Beijing Normal University, Beijing, 100875, China.
}%
\affiliation{%
Key Laboratory of Multiscale Physics, Ministry of Education, Beijing Normal University, Beijing,100875, China.
}%

\author{Pietro Gambardella}
\email{pietro.gambardella@mat.ethz.ch}
\affiliation{%
Department of Materials, ETH Zurich, Zurich 8093, Switzerland
}%

\date{\today}

\begin{abstract} 
Electrical detection of magnetization dynamics in magnetic insulators underpins both fundamental studies of magnon transport and the development of low-loss magnonic devices. In heavy-metal/magnetic-insulator heterostructures, spin pumping and spin-torque ferromagnetic resonance (ST-FMR) are widely used for this purpose and are often treated separately in different measurement geometries.
In practice, the competition between these two effects gives rise to electrical voltage signals of opposite signs, which can lead to ambiguous interpretations of the underlying physics. Here, we show how to disambiguate their respective contributions and provide a framework for interpreting experiments involving microwave excitation of magnetic insulators and detection of magnetization dynamics via spin-charge conversion in heavy metals. 
We systematically investigate spin pumping and ST-FMR in nonlocal and local devices using Pt-capped thin films of thulium and bismuth-yttrium iron garnets. We show how spin-wave character, magnetic dissipation, magnetic field orientation and device geometry govern the sign and magnitude of the resulting signals. In several cases, the voltage generated by microwave excitation changes sign between an out-of-plane or in-plane magnetic configuration. We disentangle a contribution due to spin pumping, induced by exponentially decaying propagating spin waves, and a weakly distance-dependent contribution from ST-FMR, remotely induced by inductive coupling. We show that both the spin wave excitation profile across the film thickness and magnetic damping largely determine which of the two contributions dominates. Hence, the sign of the electrical signal cannot be uniquely assigned to the chirality of the magnon modes.
Our results resolve a key ambiguity in interpreting electrical measurements of magnetization dynamics in insulators, and establish a unified framework for understanding inverse spin Hall voltage generation in magnonic devices.
\end{abstract}

\maketitle

\section{Introduction}

Magnetic insulators play a central role in emerging spintronic and magnonic technologies because they support coherent spin-wave (magnon) propagation without the need for charge transport~\cite{pirro2021advances,chumak2015magnon,demidov2017magnetization,serga2010yig,han2023coherent}. Implementing sensitive and reliable electrical readout of their magnetization dynamics is nevertheless a key requirement, both for fundamental studies of spin dynamics and for incorporating low-loss magnonic elements into practical devices and circuits~\cite{chumak2014magnon,cornelissen2015long,harder2016electrical,sheng2025control,heins2025electrical,rossi2025magnetoresistive,sandweg2011spin,lebrun2018tunable,wang2025current,schlitz2025auto}. In metal/magnetic-insulator heterostructures, several well-established techniques enable electrical readout, including spin pumping~\cite{azevedo2005dc,saitoh2006conversion,costache2006electrical,tserkovnyak2002spin,yang2018fmr}, spin-torque ferromagnetic resonance (ST-FMR)~\cite{liu2011spin,chiba2014current,kim2021theory}, and spin Seebeck effect (SSE)~\cite{uchida2008observation,uchida2010spin}.

In spin pumping, a coherently precessing magnetization injects a spin current into an adjacent metallic layer. This pure spin current is converted into a measurable dc voltage via spin-charge conversion, typically through the inverse spin Hall effect (ISHE)~\cite{sinova2015spin}. Spin pumping experiments have therefore become a central tool for characterizing magnon transport in magnetic insulators across varied frequency scales~\cite{li2020spin}. The magnitude and sign of the induced dc voltage reflect the spin Hall angle~\cite{wang2014scaling,du2014systematic,mosendz2010quantifying,mosendz2010detection,abrao2025anomalous,wang2025orbital}, the static magnetization direction~\cite{wang2025observation}, and the handedness (chirality) of the magnetization precession~\cite{liu2022switching,sheng2025control,shiota2024handedness}. 
Equally prominent is ST-FMR, in which an ac charge current in a heavy metal generates an oscillatory spin–orbit torque on an adjacent magnetic layer. When the drive frequency matches the ferromagnetic resonance, the torque excites magnetization precession. This precession modulates the device resistance, typically via spin Hall magnetoresistance (SMR)~\cite{sinova2015spin,manchon2019current} for magnetic insulators, and its mixing with the ac current produces a rectified dc voltage. ST-FMR has become a widely adopted technique for quantifying spin–orbit torque efficiencies, including in magnetic insulators~\cite{sklenar2015driving,chiba2014current,jungfleisch2017insulating}. By analyzing the amplitude and lineshape of the resonance signal, one can extract the magnitude of the current-induced torque and determine the sign of the spin Hall angle of the metallic layer~\cite{liu2011spin,mellnik2014spin,karimeddiny2021resolving}.
The SSE~\cite{bauer2012spin,kikkawa2016magnon} provides a complementary electrical probe in which a temperature gradient across a magnetic insulator generates a magnonic angular momentum current. When this angular momentum current enters the adjacent heavy metal as a spin current, it is likewise converted into a dc voltage via the ISHE. SSE measurements enable access to thermally excited magnon transport and offer an alternative route to probing spin–charge conversion and spin-current polarity in metal/magnetic-insulator heterostructures~\cite{xiao2010theory}. Microscopically, the SSE can be regarded as the incoherent version of spin pumping excited by thermal magnons~\cite{wang2025orbital}.

Although spin pumping and ST-FMR are often presented as separate effects associated with different experimental detection schemes, they naturally coexist in most devices, because both arise from microwave excitation detected by dc voltage~\cite{azevedo2011spin,chiba2014current,harder2016electrical}. 
Specifically, in an ST-FMR experiment, the radio-frequency (rf) current in the metal will inevitably generate an Oersted field that can directly drive magnetization precession in the magnetic layer. This precession pumps a spin current back into the metal and adds a spin-pumping voltage to the rectified signal. Conversely, in spin-pumping experiments, microwave fields can induce rf currents in the detector via inductive coupling. This produces ST-FMR-like rectification through spin–orbit torques. In any case, the resulting voltage is a mixture of spin-pumping and rectification signals, leading to a key interpretational difficulty~\cite{bai2013universal,obstbaum2014inverse,azevedo2011spin,okada2019spin,liu2021influence,keller2025identification,yactayo2026rf,harder2016electrical,chiba2014current,sklenar2015driving,sklenar2017unidirectional,azevedo2014competing}. The relative contributions from these two mechanisms often depend on microwave excitation field profile, device impedance, magnetic damping, electromagnetic couplings, etc., which are typically cumbersome to quantify precisely. However, assuming a secondary contribution to be negligible, when both effects are actually sizable, affects the quantification of the spin-charge conversion efficiency. Their competition may even lead to a misinterpretation of magnon chirality, since this property is commonly extracted from the sign of spin-pumping voltage directly. Beyond the quantification of these phenomena in fully metallic stacks~\cite{liu2021influence,bai2013universal,okada2019spin,sklenar2017unidirectional}, very few works have addressed the issue for magnetic insulators~\cite{sklenar2015driving}, which lets room for conflicting interpretations between experiments.

A useful comparison can be made with metallic ferromagnet/heavy-metal bilayers. In metallic systems, the rf current can flow through both the heavy metal and the ferromagnet, so the rectified voltage commonly contains contributions from anisotropic magnetoresistance (AMR), anomalous Hall effect (AHE), and SMR, while spin pumping provides an additional ISHE voltage. In magnetic insulators, by contrast, no charge current flows through the magnetic layer. This removes AMR- and AHE-based rectification inside the magnet, but does not eliminate rectification entirely, because rf currents in the adjacent heavy metal can still generate spin-orbit torques and produce SMR-based ST-FMR signals. The distinction is especially important in nonlocal geometries: in magnetic insulators, propagating spin waves can carry angular momentum without charge transport, whereas in continuous metallic ferromagnets, signals arising from rectification and eddy-current effects greatly complicate the interpretation of experiments. This distinction motivates a systematic analysis of how spin-pumping and rectification signals coexist specifically in magnetic-insulator/Pt devices, where charge transport is confined to the Pt layer, but inductive effects and spin-orbit torques can still generate competing rectification signals.

%%%%%%%%%% Note: there are several works reporting disambiguation of SP and ST-FMR signals. We are stating that this is a big problem, but are not citing studies where the big problem has been singled out. Moreover, we are not saying whether and why previous studies have been successful or unsuccessful. Consider the references below and possibly more and cite the most relevant.
% In metals:
% https://journals.aps.org/prl/abstract/10.1103/PhysRevLett.111.217602
% https://journals.aps.org/prb/abstract/10.1103/PhysRevB.89.060407
% https://pubs.aip.org/aip/jap/article/97/10/10C715/307417
% https://journals.aps.org/prb/abstract/10.1103/PhysRevB.83.144402
% https://journals.aps.org/prapplied/abstract/10.1103/PhysRevApplied.12.014040
% https://pubs.aip.org/aip/apl/article/118/13/132401/1061442
% https://pubs.acs.org/doi/full/10.1021/acs.nanolett.5c02641
% https://arxiv.org/abs/2602.14429

% In insulators:
% https://journals.aps.org/prapplied/abstract/10.1103/PhysRevApplied.2.034003
% https://journals.aps.org/prb/abstract/10.1103/PhysRevB.92.174406
% https://journals.aps.org/prb/abstract/10.1103/PhysRevB.95.224431
% https://pubs.aip.org/aip/apl/article/104/5/052402/828879
% more?

% In general:
% harder2016electrical

Here we systematically investigate how spin-pumping and spin-torque mechanisms compete in different heavy-metal/ferrimagnetic-insulator systems, and what governs the resulting dc voltage signals. We compare measurements performed in nonlocal and local devices, where Pt nanostripes are used as ISHE-voltage detectors and thin films of bismuth-yttrium or thulium iron garnets are excited by a nearby microwave antenna and direct current injection in Pt, respectively. In the nonlocal devices, we observe a clear sign reversal of the spin-pumping voltage as the magnetization is rotated from out-of-plane (OOP) to in-plane (IP). By tuning the separation distance between the microwave antenna and the Pt detector, and analyzing the frequency and angular dependence of the signals in different geometries, we separate the spin-pumping component generated by propagating spin waves and the ST-FMR component induced by inductive couplings. They are notably distinguishable by their exponential and power-law decay with antenna--detector distance, respectively. Our measurements and accompanying calculations consistently depict the competition between spin-pumping and ST-FMR contributions as the origin of the sign reversal in angular-dependent spin pumping. In low-damping films, propagating spin waves dominate the electrical response over extended distances, whereas in higher-damping systems, the inductive ST-FMR contribution can dominate even for short separations. Finally, by exciting perpendicular standing spin-wave modes, we show that non-uniform precession across the film thickness strongly suppresses spin-pumping contributions, leading to signal dominated by ST-FMR.
These findings provide practical guidelines for reliably identifying spin-transport mechanisms and for designing future magnonic and spintronic architectures based on low-damping magnetic insulators.

This paper is organized as follows. 
In Sec.~\ref{sec:theory}, we present the analytical framework describing the competition between spin pumping and ST-FMR rectification in Pt/magnetic-insulator bilayers. 
Section~\ref{sec:experiment} summarizes the sample growth, device fabrication, and measurement techniques. 
The experimental results are presented in Sec.~\ref{sec:Results}: 
Sec.~\ref{sec:nonlocalTmIG} discusses the nonlocal voltage as a function of magnetization orientation and antenna--detector distance, 
Sec.~\ref{sec:localTmIG} establishes the local ST-FMR contribution, 
Sec.~\ref{sec:spatialdecay} analyzes the spatial decay of the different mechanisms, 
Sec.~\ref{sec:damping} examines the role of magnetic damping, and 
Sec.~\ref{sec:PSSW} investigates the behavior of nonuniform spin-wave modes. 
In Sec.~\ref{sec:guidelines}, we provide practical guidelines for interpreting electrical signals in garnet/Pt heterostructures. 
Finally, Sec.~\ref{sec:Conclusions} summarizes the main conclusions of the present work.

\section{Competition Between Spin Pumping and ST-FMR rectification effects}
\label{sec:theory}

We first recall the general properties of the dc voltage generated in electrically detected ferromagnetic resonance experiments. The measured voltage can be written as
\begin{equation}
V_{\rm DC}
=
V_{\rm mix}
+
V_{\rm sp}
+
V_{\rm SSE},
\end{equation}
% where
% \begin{equation}
% V_{\rm mix}
% =
% \left\langle I_{\rm rf}(t)\delta R[\mathbf{m}(t)]\right\rangle
% \end{equation}
where $V_{\rm mix} = \left\langle I_{\rm rf}(t)\delta R[\mathbf{m}(t)]\right\rangle$ is the rectified voltage arising from the mixing between the rf current $I_{\rm rf}$ and the magnetization-dependent resistance change $\delta R[\mathbf{m}(t)]$, $V_{\rm sp}$ is the inverse-spin-Hall voltage generated by spin pumping, and $V_{\rm SSE}$ is the thermally generated spin-Seebeck contribution. The spin-pumping contribution is governed by the pumped spin current
\begin{equation}
\label{sp_geff}
\mathbf{J}_{s}^{\rm sp}
\propto
G_{\rm eff}\,
\mathbf{m}\times\dot{\mathbf{m}},
\end{equation}
where $G_{\rm eff}$ is the effective spin-mixing conductance.

In metallic ferromagnet/heavy-metal bilayers, $I_{\rm rf}$ can flow through the magnetic layer itself, so $\delta R[\mathbf{m}(t)]$ generally contains AMR, AHE, and SMR contributions. In magnetic-insulator/heavy-metal bilayers, the charge current is confined to the heavy metal, and the rectification term is primarily governed by the SMR of the Pt layer. The following analysis therefore focuses on the magnetic-insulator case, where $\delta R$ is described by the longitudinal SMR and where the competition between SMR-based ST-FMR rectification and spin pumping determines the sign of the measured voltage.

For the Pt/magnetic-insulator bilayers studied here, we follow a standard SMR-based ST-FMR analysis~\cite{kim2021theory,azevedo2011spin,chiba2014current}.
%We provide here a brief description of the spin pumping and ST-FMR rectified voltages expected in a Pt/magnetic insulator bilayer following a standard SMR-based ST-FMR analysis~\cite{kim2021theory,azevedo2011spin,chiba2014current}.
The unit magnetization vector is defined as
\begin{equation}
\mathbf{m}=
(\sin\theta\cos\varphi,\,
 \sin\theta\sin\varphi,\,
 \cos\theta)
\end{equation}
with equilibrium angles $\theta$ and $\varphi$ as shown in Fig.~\ref{fig3}. The rf current is
\begin{equation}
I_{\rm rf}(t)=(I_0\cos\omega t,0,0).
\end{equation}
Within the longitudinal SMR picture the resistance is
\begin{equation}
R_{xx}(t)=R_0+\Delta R\left[1-m_y^2(t)\right],
\end{equation}
which then is
\begin{equation}
R_{xx}(t)
=R_0+\Delta R
-\Delta R[\sin\theta(t)\sin\varphi(t)]^2 .
\end{equation}
Expanding around the equilibrium configuration gives
\begin{align}
R_{xx}(t)
&\approx R_0+\Delta R
-\Delta R\sin^2\theta\sin^2\varphi
\nonumber\\
&-\Delta R\sin^2\theta\sin2\varphi\,
\delta\varphi(t)
\nonumber\\
&-\Delta R\sin2\theta\sin^2\varphi\,
\delta\theta(t),
\end{align}
where $\Delta R>0$. The rectified ST-FMR voltage is therefore
\begin{align}
V_{\rm mix}
&=\langle I_x(t)R_{xx}(t)\rangle
\nonumber\\
&=-\Delta R I_0\sin^2\theta\sin2\varphi
\langle\delta\varphi(t)\cos\omega t\rangle
\nonumber\\
&-\Delta R I_0\sin2\theta\sin^2\varphi
\langle\delta\theta(t)\cos\omega t\rangle .
\end{align}
For the in-plane configuration $\theta=90^\circ$,
the second term vanishes and
\begin{equation}
V_{\rm mix}
=-\Delta R I_0\sin2\varphi
\langle\delta\varphi(t)\cos\omega t\rangle .
\end{equation}

We now analyze the sign of $\delta\varphi(t)$ generated by the
spin Hall torque. The spin Hall effect in Pt produces a spin current flowing
along $-\hat z$ with polarization along $\hat y$,
\begin{align}
\mathbf{J}_s
&=\frac{\theta_{\rm SH} I}{A}\times(-\hat z)
\nonumber\\
&=\hat y\,
\frac{\theta_{\rm SH} I_0}{A}\cos\omega t ,
\end{align}
assuming $\theta_{\rm SH}>0$ and $A$ as the cross-sectional area of the Pt wire. The magnetization dynamics follows the LLG equation,
\begin{equation}
\partial_t\mathbf{m}
=-\gamma\mathbf{m}\times\mathbf{B}
+\alpha\mathbf{m}\times\partial_t\mathbf{m}
+\bm{\tau}_{\rm DL},
\end{equation}
with damping-like torque
\begin{align}
\bm{\tau}_{\rm DL}
&=\gamma B_j(t)(\mathbf{m}\times\mathbf{p})\times\mathbf{m}
\nonumber\\
&\sim
\gamma B^{\rm Amp}_j\cos\omega t
(\mathbf{m}\times\hat y)\times\mathbf{m}
\nonumber\\
&\propto
\hat\varphi\,\cos\varphi\cos\omega t ,
\end{align}
where $B_j(t)$ is the damping-like torque effective field, whose amplitude $B^{\rm Amp}_j=j \hbar \theta_{\mathrm{SH}} /\left(2 e M_{\mathrm{S}} d\right)$, with $j, \hbar, \theta_{\mathrm{SH}}, e$ and $d$ being the current density, reduced Planck constant, spin Hall angle, elementary charge and layer thickness, respectively.

At resonance, the damping-like torque drives the azimuthal precession through its projection onto the $\hat\varphi$ direction. Therefore, the resonant component of the azimuthal oscillation can be written as
\begin{equation}
\delta\varphi(t)=\delta\varphi_{\rm DL}\cos\varphi\cos\omega t ,
\end{equation}
where $\delta\varphi_{\rm DL}>0$ includes the resonant susceptibility and the torque amplitude.

Consequently,
\begin{align}
V_{\rm mix}
&=
-\Delta R I_0\sin2\varphi
\left\langle\delta\varphi(t)\cos\omega t\right\rangle
\nonumber\\
&=
-\frac{1}{2}\Delta R I_0\delta\varphi_{\rm DL}
\sin2\varphi\cos\varphi
\nonumber\\
&=
-\Delta R I_0\delta\varphi_{\rm DL}
\sin\varphi\cos^2\varphi <0 .
\end{align}

In contrast, for spin pumping the emitted magnons carry spin angular momentum opposite to their equilibrium spin, i.e., along $\mathbf{m}$. The inverse spin Hall current therefore reads
\begin{align}
\mathbf{J}
&=\theta_{\rm SH}|\mathbf{J}_S|
\,\mathbf{m}\times\hat z
\nonumber\\
&=\theta_{\rm SH}|\mathbf{J}_S|
(\sin\theta\sin\varphi,
-\sin\theta\cos\varphi,
0)
\end{align}
and the measured voltage is thus
\begin{align}
V_{\rm sp}
&\sim
\theta_{\rm SH}|\mathbf{J}_S|
\sin\theta\sin\varphi
\nonumber\\
&=
\theta_{\rm SH}|\mathbf{J}_S|
\sin\varphi
>0 .
\end{align}

During ST-FMR measurements a temperature gradient between Pt and the magnetic layer can also generate a spin Seebeck signal. Since Pt is typically hotter, the spin current flows opposite to that of spin pumping, leading to~\cite{wang2024electrical}
\begin{equation}
V_{\rm SSE}<0 .
\end{equation}

\section{Experimental Details}
\label{sec:experiment}

TmIG and BiYIG thin films were grown at high temperatures by radio-frequency magnetron sputtering from $\mathrm{Tm_{3}Fe_{5}O_{12}}$ and $\mathrm{Bi_{0.8}Y_{2.2}Fe_{5}O_{12}}$ targets on (111)-oriented Y$_3$Sc$_2$Ga$_3$O$_{12}$ (YSGG) or $\mathrm{Gd_{3}Sc_{2}Ga_{3}O_{12}}$ (GSGG) substrates, as reported in Refs.~\onlinecite{wang2025ultrathin,legrand2025lattice}. 
%The substrate was mounted on a planetary holder rotating in a plane parallel to the target, positioned at a sample height of 80~mm in an off-axis configuration. Initially, the (111)-oriented Y$_3$Sc$_2$Ga$_3$O$_{12}$ (YSGG) or $\mathrm{Gd_{3}Sc_{2}Ga_{3}O_{12}}$ (GSGG) substrate was annealed for 1 hour at 750 $^\circ$C in a mixed Ar-O$_2$ atmosphere. TmIG thin films were subsequently deposited with an RF power of 30 W in a mixed Ar-O$_2$ atmosphere with gas flow rates of 100 sccm and 1 sccm, respectively, and a total pressure of 3.8 mTorr. The growth of BiYIG requires a higher O$_2$ flow rate of 5~sccm. After growth, the sample was further annealed for another 1 hour before cooldown to room temperature.
The lattice mismatch between TmIG and YSGG induces epitaxial strain, which enables perpendicular magnetic anisotropy (PMA) in TmIG. In contrast, due to the different lattice mismatch between BiYIG and YSGG or GSGG substrates, BiYIG grown on YSGG exhibits IP anisotropy, whereas BiYIG grown on GSGG exhibits PMA. In addition, the strain also modifies the magnetic dissipation. The origin of the PMA in the TmIG/YSGG and BiYIG/GSGG films is mainly assigned to tensile in-plane stress, as characterized by X-ray diffraction, detailed in Supplemental Material (Ref.~\cite{SI}).\nocite{Birkholz2005XRD,Holy1999HRXRD,Miller2022Laue,Thompson1987PseudoVoigt}

\begin{figure}
\hspace{-0.5cm}\includegraphics[width=60mm]{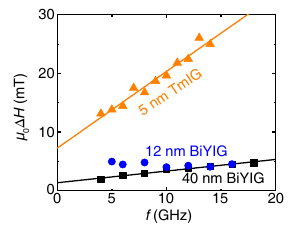}
\caption{\label{suppfig:fig1} FMR linewidth as a function of excitation frequency for the different iron garnet thin-film samples investigated in this work. Solid lines are linear fits used to extract Gilbert damping and inhomogeneous broadening. }
\end{figure}

Because the competition between spin pumping and ST-FMR heavily depends on the dissipation in the magnetic system, microwave losses were characterized as a function of frequency using broadband ferromagnetic resonance (FMR)~\cite{legrand2025implementation}. Figure~\ref{suppfig:fig1} presents the resonance linewidth of TmIG and BiYIG under excitation at different frequencies. By fitting the linewidth values to $\mu_0 \Delta H=\mu_0 \Delta H_0+2 \alpha \omega / \gamma$, a Gilbert damping $\alpha\approx$~0.0028 $\pm$ 0.0002 and an inhomogeneous broadening $\mu_0\Delta H_0\approx$~1.4 $\pm$ 0.1 mT are obtained for the 40-nm-thick BiYIG grown on YSGG; $\alpha\approx$~0.015 $\pm$ 0.001 and $\mu_0\Delta H_0\approx$~7.3 $\pm$ 0.1 mT are obtained for the 5-nm-thick TmIG grown on YSGG. For the 12-nm-thick BiYIG film grown on GSGG, the linewidth first decreases with increasing frequency and then increases again, starting from above approximately 15~GHz. As a result, a reliable damping value cannot be extracted for this sample. However, since the measurements discussed in this work are primarily performed around 5~GHz, a direct comparison of the linewidths at this frequency is sufficient. At 5~GHz, the 40-nm-thick BiYIG film exhibits the smallest dissipation, followed by the strained 12-nm-thick BiYIG film grown on GSGG, while the 5-nm-thick PMA TmIG film shows the largest dissipation. %Consistent with this trend, the sign-reversal effect is most pronounced in the TmIG sample with the highest dissipation.

For the nonlocal devices as shown in Fig.~\ref{fig1}(a), microwave antennas were patterned by electron-beam lithography. After development, a 10-nm-thick Ti adhesion layer and a 90-nm-thick Au layer were deposited by electron-beam evaporation, followed by lift-off. The Pt detector was then fabricated in a second lithography step. After development, the remaining resist was removed by an O$_2$ plasma ashing process to improve the interface quality. Subsequently, a 5-nm-thick Pt layer was deposited by dc magnetron sputtering and lifted off to form the detector nanostripe. The antenna and detector nanostripes have both a width of 760~nm and are separated by a distance $s$. For electrical detection of magnetization dynamics using the nonlocal spin-pumping technique~\cite{wang2024broad}, we apply a pulse-modulated microwave current to the Ti/Au antenna and measure the rf-induced change of voltage between the two ends of the Pt detector using a lock-in amplifier synchronized with the pulse modulation of the microwaves.

For the local devices, the Pt nanostrip was fabricated using the same lithography and dc sputtering process. In these local devices, the Pt strip had a width of 5~$\upmu$m and a length of 150~$\upmu$m. Electrical contacts were then defined in a separate lithography step, followed by electron-beam evaporation of Ti/Au and lift-off. In the ST-FMR measurements, current injection and voltage readout are performed through a single Pt nanostripe using a bias tee, which separates the rf and low-frequency paths. The pulse-modulated rf signal from a microwave source is applied via the rf port, driving magnetization dynamics through spin–orbit torques. The resulting rectified voltage is extracted from the low-frequency port of the bias tee and measured using a lock-in amplifier.

\section{Results and Discussion}
\label{sec:Results}
\subsection{Nonlocal voltage dependence on magnetization orientation and antenna-detector distance}
\label{sec:nonlocalTmIG}

\begin{figure}
\hspace{-0.5cm}\includegraphics[width=90mm]{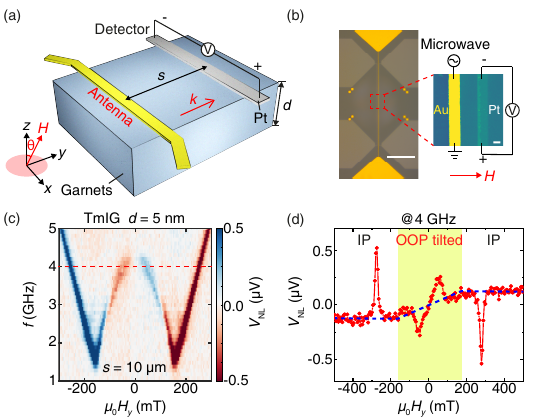}
\caption{(a) Schematic illustration of a nonlocal device fabricated on a garnet thin film. 
(b) Optical microscope image and atomic force microscopy (AFM) topographic map of a 5-nm-thick TmIG device, showing the microwave antenna and the Pt detector. The scale bar in optical image is 50~$\upmu$m and in AFM image is 500~nm. 
(c) Field-dependent nonlocal voltage spectra measured from a Pt nanostripe located $10~\upmu\mathrm{m}$ away from the microwave antenna on a 5-nm-thick TmIG film with PMA.
The external magnetic field is applied in the film plane along $y$.
(d) Linecut at 4~GHz extracted from (c). The yellow background indicates the range of fields over which the magnetization is not saturated along $y$, the blue dashed line models the spin Seebeck effect.
}
\label{fig1}
\end{figure}

\begin{figure*}
\hspace{-0.5cm}\includegraphics[width=175mm]{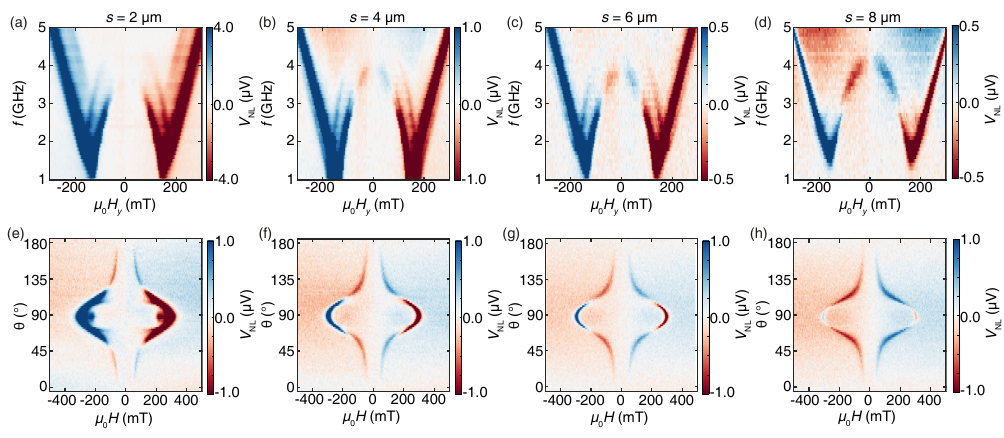}
\caption{(a-d) Field-dependent nonlocal voltage spectra measured on 5-nm-thick TmIG devices with antenna--detector separation distances ranging from 2 to 8~$\upmu$m in steps of 2~$\upmu$m. 
(e-h) Polar-angle-dependent measurements performed at a fixed excitation frequency of 5~GHz. For all panels, the microwave excitation power is +15~dBm.
}
\label{fig2}
\end{figure*}

We first study a 5-nm-thick epitaxial film of TmIG in the nonlocal geometry shown in Figs.~\ref{fig1}(a) and~\ref{fig1}(b), where the signal is commonly interpreted as spin pumping. The applied microwave current generates a local oscillating magnetic field around the antenna that excites coherent propagating spin waves in the magnetic layer. The excited spin waves propagate to the detector, where they generate a spin current via spin pumping, subsequently converted into a charge current and a voltage through the ISHE. In our convention, this nonlocal spin-pumping voltage is expected to be negative for Pt and magnetic field along $y$.

Figure~\ref{fig1}(c) shows the nonlocal voltage ($V_{\mathrm{NL}}$) spectra resulting from microwave excitation at an antenna--detector separation $s = 10~\upmu$m, for different fields and frequencies. An IP external magnetic field is applied along the $y$ axis ($\theta=90^\circ$), and the microwave power is fixed at $+15$~dBm. These spectra are dominated by the ferromagnetic resonance (FMR) mode of the TmIG thin film, which exhibits a gradual magnetic reorientation from OOP to fully IP around 150~mT, consistent with the PMA in our TmIG film. Above this reorientation field, the resonance frequency increases monotonically, following the conventional Kittel relation. 
The in-plane-field-induced reorientation inferred from the spin-wave spectra was further corroborated by static magnetic characterization of the TmIG(5~nm)/YSGG film by superconducting quantum interference device (SQUID) magnetometry, as shown in Ref.~\onlinecite{SI}. 
The measured in-plane loop shows a hard-axis-like behavior with gradual saturation, confirming the strength of the PMA in the film.

Unexpectedly, at lower magnetic fields where the magnetization is still tilted, the nonlocal voltage has the opposite sign compared to the signal for IP magnetic orientation. 
We show a linecut at 4~GHz in Fig.~\ref{fig1}(d), which highlights the anomalous sign change between the two FMR peaks. As shown in Appendix~\ref{app:simulations}, the possibility that a reversed spin-wave handedness arises from tilted magnetization is excluded based on micromagnetic simulations. The observed sign change therefore indicates the presence of an additional mechanism beyond spin pumping, which contributes a voltage of opposite sign. It is worth noting that the SSE only generates a smooth background signal to the nonlocal voltage, compared to the peaked signals at resonance, as indicated by the blue dashed line in Fig.~\ref{fig1}(d).

To gain further insight into this anomaly, we investigate a series of devices with separation distance $s$ ranging from 2 to 8~$\upmu$m. The field-dependent nonlocal voltage spectra measured in these devices are shown in Figs.~\ref{fig2}(a-d). As $s$ decreases, the sign reversal over magnetic reorientation is gradually suppressed and becomes nearly invisible for $s = 2~\upmu$m. This observation further excludes the possibility that the sign reversal originates from an intrinsic left-handed mode, which should be preserved over spin-wave propagation. Note that in the devices with small $s$, two distinct resonance modes are observed, especially for magnetic fields below IP saturation. They correspond to the co-excitation of uncapped and Pt-capped TmIG regions. The Pt capping layer generates additional IP magnetic anisotropy due to an interfacial mechanism, as established in previous studies~\cite{lee2020interfacial,lee2023large,wang2025current}. Hence, the PMA in TmIG is locally reduced, which results in the lower (higher) frequency mode below (above) IP saturation. 

To corroborate this observation, we fix the excitation frequency at 5~GHz and perform measurements while rotating the magnetic-field polar angle. The choice of this frequency above the zero-field resonance ensures that, at the corresponding IP and OOP resonance fields, the magnetic field is strong enough to saturate the magnetization, with a continuous evolution of the magnetization polar angle in-between. The measurement of devices with different separation distances $s$ enables us to find the conditions for sign reversal in varied experimental configurations. As shown in Figs.~\ref{fig2}(e-h), a sign reversal systematically occurs as the magnetization evolves towards the OOP easy axis. Consistently with Figs.~\ref{fig2}(a-d), the polar angle at which a sign reversal occurs in Figs.~\ref{fig2}(e-h) reduces for devices with larger $s$. This indicates that the anomalous contribution predominates further away from the antenna. In other words, the spatial decay of this additional mechanism must be much slower than the attenuation of the propagating spin waves inducing spin pumping, which are expected to decay exponentially with $s$. Frequency-dependent polar-angle measurements on the $s = 6~\upmu$m device in Appendix~\ref{app:freqdep} reveal that the shortening of the propagation length at higher excitation frequencies further attenuates propagating spin waves and broadens the angular range dominated by the anomalous contribution. We show below that this anomalous contribution originates from ST-FMR induced by inductive coupling between the antenna and the Pt detector. 

According to Figs.~\ref{fig2}(a-d), the sign reversal at low fields occurs predominantly in the Pt-capped mode. This observation indicates that the additional mechanism leading to a signal of opposite sign is closely linked to the TmIG region beneath the Pt detector.

\subsection{Local voltage dependence on magnetization orientation}
\label{sec:localTmIG}
\begin{figure}
\hspace{-0.5cm}\includegraphics[width=90mm]{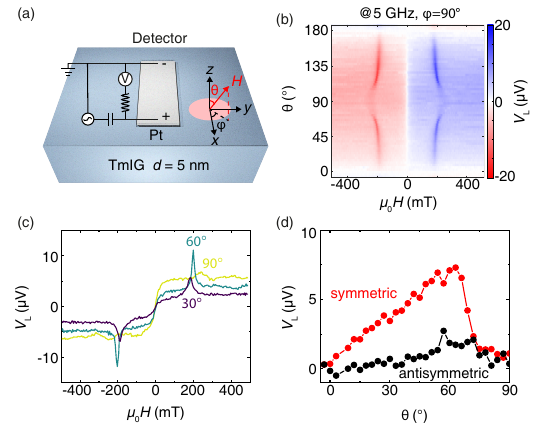}
\caption{Local voltage detection in the 5-nm-thick TmIG film. (a) Schematic illustration of the ST-FMR device geometry and direction of the applied magnetic field.  
(b) Polar-angle-dependent local voltage spectra.  
(c) Linecuts extracted from (b) at polar angles of $30^\circ$, $60^\circ$, and $90^\circ$.  
(d) Extracted symmetric and antisymmetric Lorentzian components in the ST-FMR resonance peak as a function of polar angle.
}
\label{fig3}
\end{figure}

To directly verify the ST-FMR origin of the anomalous contribution at the detector reported in Sect.~\ref{sec:nonlocalTmIG}, we employ a local device fabricated on the same chip, applying the magnetic field in the $yz$ plane with $\varphi = 90^\circ$, as shown in Fig.~\ref{fig3}(a). The resulting polar-angle-dependent local voltage ($V_{\rm{L}}$) spectra are presented in Fig.~\ref{fig3}(b). 
This voltage exhibits the same positive sign over the entire polar-angle range with $\mu_0H_y>0$. This sign is consistent with that observed in the long-distance nonlocal spin-pumping devices when the contribution from propagating spin waves is strongly attenuated, confirming that the additional mechanism arises from local ST-FMR at the Pt detector. 
Since the resistance change due to SMR is given by $\Delta R_{\mathrm{SMR}}\sin^{2}\theta\sin^{2}\varphi$, the symmetric and antisymmetric ST-FMR components follow a $\sin 2\varphi \cos \varphi$ dependence when the magnetic field is applied in the $xy$ plane~\cite{mellnik2014spin,harder2016electrical} (Sect.~\ref{sec:theory}). As a result, no ST-FMR signal is expected for $\varphi =90^\circ$ when the field is applied in the plane ($\theta = 90^\circ$), which is consistent with the linecuts in Fig.~\ref{fig3}(c), extracted from Fig.~\ref{fig3}(b). Peaks from these linecuts can be decomposed into symmetric and anti-symmetric Lorentzian components with respect to $H_{\text{res}}$, as summarized in Fig.~\ref{fig3}(d). Only when the field tilts the magnetization partially OOP does a finite ST-FMR signal appear. Section~\ref{sec:theory} provides detailed analytical calculations showing that the signs of the spin-pumping and ST-FMR contributions are intrinsically opposite.

In both the nonlocal [Fig.~\ref{fig1}(d)] and local voltage spectra [Fig.~\ref{fig3}(c)], a background signal from the SSE is observed, arising from a temperature gradient established by microwave-induced on-chip Joule heating. This SSE background exhibits the same sign in both types of measurements. From the ST-FMR experiment, it is clear that the Pt layer is hotter than the TmIG, since the microwave current is directly injected into the Pt. This observation indicates that in the nonlocal spin-pumping geometry, although the microwave excitation is applied nonlocally through the antenna, the Pt detector can still become hotter than the TmIG layer~\cite{shan2016influence}. This heating is caused by the rf currents that are inductively generated in the Pt detector. We note that the sign of the observed SSE configuration is in agreement with literature for YIG/Pt when the Pt is hotter~\cite{schreier2015sign}.

\subsection{Spatial decay of nonlocal and local signals}
\label{sec:spatialdecay}
To quantitatively analyze the spatial decay of the conventional propagating spin-wave and inductive-coupling-induced ST-FMR contributions, we extract the nonlocal voltages as a function of polar angle by performing symmetric Lorentzian fits to the measured spectra. As an example, the results for the device with a separation distance $s=$ 10~$\upmu$m are shown in Fig.~\ref{fig4}(a), where the contributions from ST-FMR and spin pumping remain in the linear regime due to the relatively long antenna–detector distance. 
For $\theta = 90^\circ$, the nonlocal voltage arises predominantly from propagating spin waves, because the SMR-based ST-FMR contribution is symmetry-suppressed, as seen in Fig.~\ref{fig3}. 
We therefore use the voltage at $\theta = 90^\circ$ as a reference and approximate the angular dependence of the propagating spin-wave spin-pumping contribution by
\begin{equation}
V_{\rm sp}(\theta)
=
V_{\rm sp}(90^\circ)\sin^3\theta ,
\label{eq:sp_sw_sin3}
\end{equation}
where $\sin^3\theta$ scaling contains one factor of $\sin\theta$ from the ISHE detection efficiency and an additional $\sin^2\theta$ factor from the antenna excitation efficiency, which is proportional to the microwave absorption and therefore to the square of the transverse rf torque. 
This approximation gives the blue curve in Fig.~\ref{fig4}(a).

Within this framework, the ST-FMR signal can be roughly estimated by subtracting the $\sin^3\theta$ spin-pumping contribution from the raw data. 
The inferred ST-FMR component, shown as red symbols in Fig.~\ref{fig4}(a), reproduces the overall angular trend measured independently in the local device, shown in Fig.~\ref{fig3}(d).
Beyond this approximate estimate, full quantitative agreement is hard to achieve because the actual angular dependence of nonlocal spin pumping is also be affected by the propagation itself: spin-wave dispersion, group velocity, propagation length, damping, mode ellipticity, and $k$-dependent antenna excitation efficiency. These corrections are not expected to significantly alter the overall angular dependence and are therefore not considered further.
Instead, the results for devices with different separation distances are summarized in Fig.~\ref{fig4}(b). 
Notably, the data measured for $s=2~\upmu$m deviates from the smooth angular dependence observed at larger distances, indicating nonlinear magnonic contributions~\cite{wang2023deeply}.

%For $\theta = 90^\circ$, the nonlocal voltage arises solely from propagating spin waves, as discussed in Fig.~\ref{fig3}. Using that both the ISHE detection efficiency and the antenna excitation efficiency incorporate a $\sin\theta$ dependence, the angular dependence of the spin-pumping contribution can be approximated by a $\sin^2\theta$ term. Using the voltage at $\theta = 90^\circ$ as a reference, we construct the expected angular dependence of the propagating spin-wave spin-pumping contribution, shown as the blue curve in Fig.~\ref{fig4}(a). We note that the actual angular dependence of spin pumping may be more complex due to additional factors, such as polar-angle-dependent changes in magnon dispersion. Within this simplified framework, the ST-FMR contribution can be roughly estimated by subtracting the expected spin-pumping signal from the raw data. The inferred ST-FMR component, shown as red symbols in Fig.~\ref{fig4}(a), exhibits an angular dependence consistent with the one measured independently in the local device, shown in Fig.~\ref{fig3}(d). Because the ST-FMR contribution depends overall on many factors, we do not aim to describe it by an analytical expression. Instead, the results for devices with different separation distances are summarized in Fig.~\ref{fig4}(b). Notably, the data measured for $s=2~\upmu$m deviates from the smooth angular dependence observed at larger distances, indicating  nonlinear magnonic contributions~\cite{wang2023deeply}.

The amplitude of the negative-sign contribution due to propagating spin waves decays much more rapidly with distance than the positive-sign contribution from ST-FMR. The slower spatial dependence of nonlocal ST-FMR can be attributed to the $1/s$ decay of the associated electromagnetic fields, in contrast to the exponential decay affecting propagating spin waves. This idea is confirmed by analyzing the spin-pumping contribution, directly obtained from $\theta = 90^\circ$ where ST-FMR vanishes. As shown in Fig.~\ref{fig4}(c), the evolution of the spin-pumping contribution with $s$ is well described by an exponential decay, consistent with the expected propagative damping for spin waves. A significant deviation occurs for $s=2~\upmu$m (red arrow), consistent with the nonlinear magnonic contributions identified above. From the fit, we obtain a decay length of $2.4 \pm 0.8~\upmu$m at 5~GHz. In contrast, the ST-FMR contribution is well described by a $1/s$ dependence [Fig.~\ref{fig4}(d)], as expected for an inductive coupling mechanism.

We also considered a possible remote spin-pumping contribution, due to the inductively generated microwave electromagnetic field in the Pt detector directly exciting magnetic precession underneath. Such a contribution would have the same ISHE polarity as spin pumping generated by propagating spin waves, because the sign is determined by $\mathbf{m}\times\dot{\mathbf{m}}$ and the equilibrium magnetization direction. Therefore, this mechanism cannot account for the slowly decaying voltage component, which has the opposite sign to the propagating-spin-wave spin-pumping contribution and is instead attributed to inductive-current-induced ST-FMR in the Pt detector. Moreover, in the in-plane configuration $\theta=90^\circ$ and $\varphi=90^\circ$, SMR-based rectification is suppressed by symmetry. A sizable electromagnetic-field-induced local spin-pumping signal would then appear as a slowly decaying background in the distance dependence. Instead, the data in Fig.~\ref{fig4}(c) are well described by an exponential decay without a detectable power-law tail. We therefore conclude that direct electromagnetic-field-induced spin pumping at the detector is negligible within our experimental sensitivity, and that the slowly decaying opposite-sign contribution is dominated by inductively generated ST-FMR rectification.

\begin{figure}
\hspace{-0.5cm}\includegraphics[width=90mm]{fig4_v7.pdf}
\caption{Analysis of nonlocal detection for the 5-nm-thick TmIG film.  
(a) Peak voltage at resonance as a function of magnetic field polar angle, for $s=10~\upmu$m (black symbols). 
The blue curve represents the estimated propagating spin-wave spin-pumping contribution using a phenomenological $\sin^3\theta$ angular envelope normalized to the signal at $\theta=90^\circ$. 
The red symbols show the difference between the raw data and this spin-pumping estimate, highlighting the ST-FMR contribution.
(b) Extracted peak voltage at resonance as a function of magnetic field polar angle, for devices with different $s$.  
(c) Pure spin-pumping contribution at $90^\circ$, as a function of $s$. The solid curve represents an exponential fit, shown on a logarithmic scale. The red arrow indicates that nonlinear effects influence the device with $s=$ 2~$\upmu$m.  
(d) ST-FMR contribution, averaged over the polar-angle range from $15^\circ$ to $30^\circ$, where the propagating spin-wave contribution is strongly suppressed. The solid line is a linear fit to $1/s$, indicating a power-law decay with antenna–detector distance.
}
\label{fig4}
\end{figure}

\subsection{Influence of damping on nonlocal and local signals}
\label{sec:damping}

\begin{figure}
\hspace{-0.5cm}\includegraphics[width=90mm]{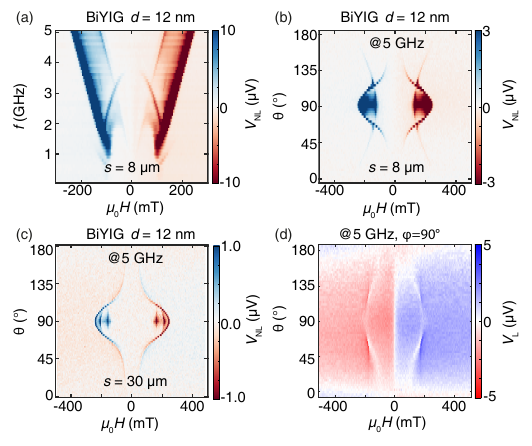}
\caption{\label{suppfig:fig3}(a) IP field-dependent nonlocal voltage spectra measured on a 12-nm-thick BiYIG film grown on a GSGG substrate, with an antenna--detector separation distance $s$ of $8~\upmu\mathrm{m}$.
(b,c) Polar-angle-dependent nonlocal voltage measurements performed at a fixed excitation frequency of 5~GHz on devices with antenna--detector separations of $8~\upmu\mathrm{m}$ (b) and $30~\upmu\mathrm{m}$ (c).
(d) Polar-angle-dependent local voltage spectra measured on the 12-nm-thick BiYIG film grown on a GSGG substrate. }
\end{figure}

Because the intrinsic magnetic dissipation in TmIG is relatively large, propagating spin waves decay rapidly over distance and the ST-FMR contribution easily dominates the nonlocal voltage. We therefore turn to a 12-nm-thick BiYIG grown on GSGG substrate, which exhibits lower damping and a considerably longer propagation length for spin waves. As shown in Fig.~\ref{suppfig:fig3}(a), the IP-field-dependent nonlocal voltage spectra appear very similar to that observed in the TmIG film, including the OOP to IP transition and the presence of uncapped and Pt-capped modes. 
The corresponding polar-angle-dependent nonlocal voltage spectra measured at an excitation frequency of 5~GHz are shown in Fig.~\ref{suppfig:fig3}(b). In contrast to the TmIG device with the same separation distance, no sign reversal is observed in the BiYIG device. These results indicate that propagating spin waves play a more dominant role in BiYIG due to their longer decay length. It is also worth noting that the uncapped mode has $M_{\rm eff}=M_{\rm s}+H_{\rm ani}<0$, whereas the Pt-capped mode has $M_{\rm eff}>0$. As a result, their resonance positions shift in opposite directions when the magnetic field is rotated from IP to OOP.

To reveal the contribution from inductive coupling in the 12-nm-thick PMA BiYIG film, the separation distance $s$ has to be increased until the propagating spin waves are largely attenuated and their voltage contribution becomes smaller than that from ST-FMR generated inductively. As shown in Fig.~\ref{suppfig:fig3}(c), when $s$ is increased to $30~\upmu$m, a clear sign reversal of the nonlocal voltage is observed again at large polar angles. The ST-FMR spectra shown in Fig.~\ref{suppfig:fig3}(d) further confirm that the ST-FMR and propagating spin-wave contributions have opposite signs. Compared with the ST-FMR spectra measured in the TmIG film, the antisymmetric component is more pronounced in 12-nm-thick PMA BiYIG, indicating a stronger field-like torque contribution~\cite{harder2016electrical}.

We next study a system with even lower dissipation by using a 40-nm-thick BiYIG film grown on a YSGG substrate, with magnetic dissipation characterized by a Gilbert damping $\alpha\approx$~0.0028 $\pm$ 0.0002 and an inhomogeneous broadening $\mu_0\Delta H_0\approx$~1.4 $\pm$ 0.1 mT (Sect.~\ref{sec:experiment}).
First, Fig.~\ref{fig5}(a) shows the polar-angle-dependent voltage spectra measured in a local device at 5~GHz, under the same conditions as used above for the TmIG film. The sign of the SSE background is consistent with that observed in TmIG [Fig.~\ref{fig3}(b)] and 12-nm-thick BiYIG [Fig.~\ref{suppfig:fig3}(d)]. However, the sign of the resonance peak is reversed, as shown in Fig.~\ref{fig5}(b). 
This reversal indicates that, in the 40-nm-thick BiYIG/YSGG device, spin pumping becomes the dominant contribution even in the local measurement geometry. 
We emphasize that the relative weights of spin pumping and SMR-based ST-FMR rectification are not determined by magnetic damping alone. 
Spin pumping depends on the effective spin-mixing conductance $G_{\rm eff}$ at the Pt/garnet interface (Eq.~\ref{sp_geff}), whereas the rectified ST-FMR voltage scales with the SMR ratio of the Pt strip. 
To assess whether the dominance of spin pumping in this device could simply originate from an anomalously weak SMR, we performed angle-dependent magnetoresistance measurements on the same BiYIG(40 nm)/YSGG device, as described in Ref.~\onlinecite{SI}. 
The extracted average SMR ratio is
$\Delta R_{\rm SMR}/R=(2.14\pm0.08)\times10^{-4}$. This value is comparable to the commonly reported $10^{-4}$-level SMR ratios in YIG/Pt~\cite{Nakayama2013SMR,Vlietstra2013SMR,Althammer2013SMR,Puetter2017SMR,Reustlen2025SMR}, although it is smaller than the largest values obtained for optimized in-situ YIG/Pt interfaces. 
This suggests that the predominance of spin pumping in the local geometry cannot be solely attributed to a suppressed SMR.

The low magnetic dissipation of the 40-nm-thick BiYIG film provides a clear explanation for the enhanced local spin-pumping contribution. 
For a given microwave Oersted field or spin-orbit torque, reduced damping increases the resonant dynamical susceptibility and therefore the precession amplitude. 
Since the spin-pumping current is generated by $\mathbf{m}\times\dot{\mathbf{m}}$, the enhanced magnetization dynamics can strongly increase the inverse-spin-Hall voltage. 
In comparison, the SMR-based ST-FMR rectification voltage scales only linearly with the dynamic magnetization amplitude, because it arises from the mixing between the rf current and the resistance oscillation induced by the magnetization precession. Thus, for a fixed microwave drive, reducing the damping enhances the spin-pumping signal approximately as $|\delta \mathbf{m}|^2$, whereas the ST-FMR rectification increases only as $|\delta \mathbf{m}|$.
We therefore interpret the local spin-pumping dominance in BiYIG(40 nm)/YSGG as resulting from the combined effect of a finite Pt/garnet spin transparency and the enhanced dynamical susceptibility associated with low magnetic damping. 
A more quantitative identification of the roles of $G_{\rm eff}$, SMR ratio, microwave current distribution, and magnetic damping would require a dedicated series of samples with independently controlled interface transparency and dissipation, which is beyond the scope of this work. Only when the polar angle approaches $\theta=0^\circ$ or $180^\circ$ does the resonance peak become anti-symmetric or show a sign reversal, as shown in Fig.~\ref{fig5}(b). This behavior highlights that the intrinsic ST-FMR contribution dominates only under the condition that the local or nonlocal spin pumping is strongly attenuated.

\subsection{Nonlocal and local signals from nonuniform spin wave modes}
\label{sec:PSSW}

\begin{figure}
\hspace{-0.5cm}\includegraphics[width=90mm]{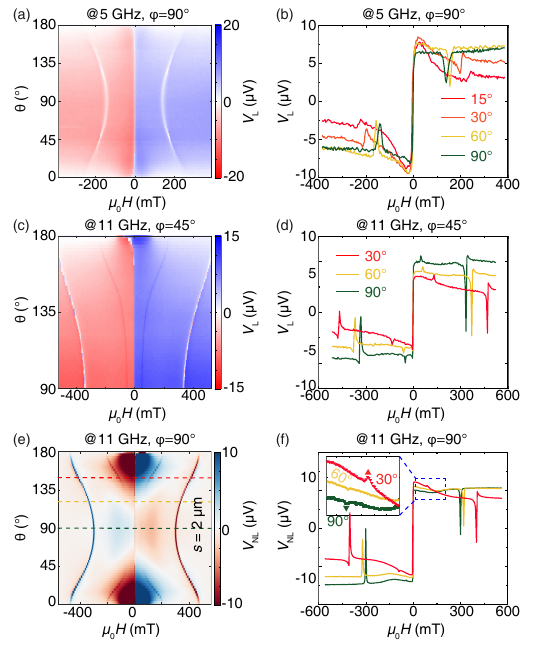}
\caption{Local and nonlocal detection for the 40-nm-thick BiYIG film. (a) Polar-angle-dependent local device voltage spectra measured at 5~GHz and $\varphi=90^\circ$. 
(b) Representative linecuts from (a) at selected polar angles, showing that the FMR peak mainly arises from spin pumping. 
(c) Polar-angle-dependent local device voltage spectra measured at 11~GHz and $\varphi=45^\circ$, where the PSSW mode becomes accessible together with FMR. 
(d) Representative linecuts from (c) at selected polar angles, highlighting that the FMR and PSSW peaks have opposite signs. 
(e) Polar-angle-dependent nonlocal device voltage spectra for antenna--detector separation $s=2~\upmu\mathrm{m}$ (+15~dBm power). The SSE background is subtracted to provide a clearer observation of the PSSW mode.
(f) Linecuts from (e) at different polar angles, illustrating the distinct signatures of the FMR and PSSW peaks.
}
\label{fig5}
\end{figure}

Perpendicular standing spin-wave (PSSW) modes can be easily excited by nonuniform magnetic fields in films of relatively large thickness. By increasing the excitation frequency to 11~GHz and setting the azimuthal angle to $\varphi = 45^\circ$ to maximize signal from the PSSW, we can excite and detect the first-order PSSW mode in the 40-nm-thick BiYIG film. Figure~\ref{fig5}(c) shows polar-angle-dependent measurement of the same local device used for the measurements reported in Fig.~\ref{fig5}(a,b). Two distinct modes are observed: the low-field mode corresponds to the PSSW, while the high-field mode is the FMR. From linecuts extracted at different polar angles, shown in Fig.~\ref{fig5}(d), the sign of the FMR peak is consistent with that at 5~GHz, indicating that it is dominated by spin pumping also at $\varphi = 45^\circ$. In contrast, the sign of the PSSW peak corresponds to ST-FMR. This indicates that for the PSSW mode, the ST-FMR contribution dominates, while the spin-pumping contribution is considerably attenuated. 
This observation can be rationalized by considering the less efficient Oersted field excitation of the PSSW mode compared to the FMR mode, since the mode overlap between the inhomogeneous magnetization dynamics of the PSSW and the antenna field is significantly smaller.

The dominant role of ST-FMR for the PSSW mode is also observed in the nonlocal configuration. Figure~\ref{fig5}(e) shows measurements of the polar-angle-dependent nonlocal voltage spectra at 11~GHz on a device with $s = 2~\upmu$m, for which both FMR and PSSW modes are present. The SSE background is subtracted in Fig.~\ref{fig5}(e) to provide a clearer observation of the PSSW mode. The FMR (high-field) mode exhibits a consistent sign over the entire polar-angle range, indicating that the contribution from propagating spin waves remains dominant. In contrast, the PSSW (low-field) mode shows a more complex behavior. Linecuts extracted at several polar angles are shown in Fig.~\ref{fig5}(f). In the $\theta=90^\circ$ (IP) configuration, the sign of the PSSW peak matches that of the FMR, and is dominated by spin pumping. As the polar angle increases, the PSSW peak reduces, nearly vanishes around $\theta=60^\circ$, before reversing sign, as seen for $\theta=30^\circ$ where the ST-FMR contribution becomes dominant. This behavior resembles that of the FMR seen in Figs.~\ref{fig5}(a,b), but the ST-FMR-dominated polar angle region extends closer towards $\theta=90^\circ$ since the Oersted field excitation of PSSWs is less effective compared to the spin-orbit torque excitation, as discussed above. We have further investigated nonlocal detection of PSSW modes in TmIG films with different thicknesses. Due to the larger magnetic dissipation further suppressing the spin pumping contribution, the opposite signs for nonlocal detection of PSSW and FMR modes becomes more pronounced. These measurements are presented in Appendix~\ref{app:PSSW}.

\begin{table*}
\caption{\label{tab:nonlocal_summary}Summary of local and non-local detection results (various antenna--detector distances $s$) obtained for the different iron garnet film compositions, thicknesses, magnetic configurations, and spin-wave modes investigated in this work. The signs refer to external fields applied with positive projection along $y$. Here, ``$+$'' indicates a clear signal with positive sign, ``$-$'' a clear signal with negative sign, ``0'' a negligible or undetected signal, and ``n/a'' not measured.}
\squeezetable
\small
\setlength{\tabcolsep}{3pt}
\begin{ruledtabular}
\begin{tabular}{llcccccccc}
Composition & Configuration (mode) & \multicolumn{1}{c}{Linewidth (mT)} &
\multicolumn{1}{c}{Local} &
\multicolumn{6}{c}{Non-local ($s$ in $\upmu$m)} \\
(thickness) & & \multicolumn{1}{c}{(@ Frequency)} &
 &
\multicolumn{1}{c}{$s=2$} &
\multicolumn{1}{c}{$s=4$} &
\multicolumn{1}{c}{$s=6$} &
\multicolumn{1}{c}{$s=8$} &
\multicolumn{1}{c}{$s=10$} &
\multicolumn{1}{c}{$s=30$} \\
\hline
TmIG (5 nm)   & IP (FMR)        & 13.8 (@5~GHz)   & $0$ & $-$ & $-$ & $-$ & $-$ & $-$ & n/a \\
TmIG (5 nm)   & OOP tilted (FMR) & n/a           & $+$ & $+$ & $+$ & $+$ & $+$ & $+$ & n/a \\
TmIG (36 nm)   & IP (PSSW) & n/a           & n/a & $0$ & n/a & n/a & n/a & n/a & n/a \\
TmIG (36 nm)   & OOP tilted (PSSW) & n/a           & n/a & $+$ & n/a & n/a & n/a & n/a & n/a \\
TmIG (72 nm)   & IP (PSSW) & n/a           & n/a & $-$ & n/a & n/a & n/a & n/a & n/a \\
TmIG (72 nm)   & OOP tilted (PSSW) & n/a           & n/a & $+$ & n/a & n/a & n/a & n/a & n/a \\
\hline
BiYIG (12 nm) & IP (FMR)        & 5.0 (@5~GHz)    & $0$ & n/a & n/a & n/a & $-$ & n/a & $-$ \\
BiYIG (12 nm) & OOP tilted (FMR) & n/a           & $+$ & n/a & n/a & n/a & $-$ & n/a & $+$ \\
BiYIG (40 nm) & IP (FMR)               & 2.3 (@5~GHz)     & $-$ & $-$ & n/a & n/a & n/a & n/a & n/a \\
BiYIG (40 nm) & OOP tilted (FMR)       & n/a            & $-$ & $-$ & n/a & n/a & n/a & n/a & n/a \\
BiYIG (40 nm) & IP (PSSW)         & 3.5 (@11~GHz)    & $+$ & $-$ & n/a & n/a & n/a & n/a & n/a \\
BiYIG (40 nm) & OOP tilted (PSSW) & n/a            & $+$ & $+$ & n/a & n/a & n/a & n/a & n/a \\
\end{tabular}
\end{ruledtabular}
\end{table*}

\section{Practical guidelines for interpreting electrical signals in garnet/Pt devices}
\label{sec:guidelines}

We now summarize the varied experimental situations encountered above into practical guidelines, aiming to minimize misinterpretation of electrically detected magnetization dynamics in heavy-metal/ferrimagnetic-insulator heterostructures, where spin pumping, ST-FMR rectification, and a spin-Seebeck background often coexist. A qualitative overview of the regimes observed in this work is provided in Table~\ref{tab:nonlocal_summary}, which can be used as a quick reference for what type of signal can be expected for a given material dissipation, mode character, magnetic configuration, and antenna--detector separation.

\paragraph*{(A) Use angular dependences to separate mechanisms (polar and azimuthal sweeps).}
Angular scans provide strong constraints because spin pumping and rectification have distinct selection rules.
For example, the voltage induced by nonlocal ST-FMR depends on the relative orientation between current and magnetization through the magnetoresistance, and can vanish for specific in-plane azimuthal angles (depending on whether AMR-type or SMR-type contributions dominate). In contrast, spin pumping detected by the ISHE primarily follows from the equilibrium magnetization direction and the efficiency of excitation/detection. In simple geometries, it is often well approximated by a smooth function of polar angle (e.g., a $\sin^2\theta$ trend). Practically:
(i) combine polar-angle sweeps with at least one azimuthal control angle chosen to suppress rectification;
(ii) if the sign persists across conditions where rectification is symmetry-suppressed, it is more likely to be spin pumping;
(iii) if the polarity matches that of a local ST-FMR device fabricated on the same chip, inductive ST-FMR is implicated.

\paragraph*{(B) Compare spin-wave propagation length and antenna--detector separation.}
Which of propagating-spin-wave spin pumping or inductive ST-FMR dominates in the signal critically depends on dissipation. To ensure a regime where spin pumping is unaffected by ST-FMR, antenna--detector separation must be kept much below the spin-wave decay length, and inductive rf fields need to be screened. Measuring the broadband FMR linewidth
$\Delta{}H(f)$ to extract damping and inhomogeneous linewidth, and using the magnon dispersion relation to obtain the group velocity, provides the spin-wave decay length. The inductive rf fields are more efficiently screened by (multiple-meander) coplanar waveguide antenna designs, at the price of wavevector bandwidth. Measurements at lower frequencies are preferable, since inductive coupling increases with frequency and the magnon decay length is larger at small frequencies. This also yields a practical test as a means of verification: by varying $s$ on otherwise identical devices, a clear exponential attenuation indicates a propagating-spin-wave contribution from spin pumping, while a slower power-law attenuation reveals inductive ST-FMR contributions.

\paragraph*{(C) Identify which spin-wave mode is probed (no assignment of chirality from voltage sign alone).}
Electrical detection spectra in nanodevices frequently contain multiple resonance modes, e.g., uncapped vs Pt-capped regions, or quasi-uniform FMR vs. PSSW modes. Since these different eigenmodes have different dynamical magnetization profiles (across the film thickness and laterally), they show different efficiencies for excitation by the antenna rf field and detection by the current-induced torques in Pt. The sign of the resulting electrical signal does not have to be the same for all modes. In practice:
(i) classify each peak by comparison with the Kittel relation and tracking thickness-dependent shifts (PSSW);
(ii) when two close modes coexist (e.g., for capped/uncapped layers), use geometry-dependent trends such as visibility against antenna--detector separation $s$ and/or micromagnetics to identify each branch.
As illustrated by Table~\ref{tab:nonlocal_summary}, PSSW modes may exhibit an opposite signal compared to FMR simply because the antenna less efficiently excites the corresponding propagating spin waves. 

\paragraph*{(D) Inspect lineshape and background to avoid mixing resonant and nonresonant contributions.}
In heavy-metal/magnetic-insulator heterostructures, a genuine spin-pumping voltage detected by ISHE yields a predominantly \emph{symmetric} Lorentzian lineshape in field sweeps.
A pronounced \emph{antisymmetric} component is instead a strong indicator of a rectification contribution due to ST-FMR. However, the symmetric part can also contain rectification contributions, even in magnetic insulators.

%\paragraph*{(F) Implement minimal control experiments to pin down inductive rectification.}
%Because inductive coupling is a circuit-mediated effect, it can often be identified (or reduced) by electrical controls at the detector:
%(i) compare open-circuit vs well-defined load impedance at the Pt detector;
%(ii) modify bonding/wiring geometry to alter pickup;
%(iii) benchmark with a ``local'' ST-FMR device (as in Fig.~\ref{fig3}) to establish the intrinsic rectification polarity for the same stack.
%Agreement between the nonlocal ``anomalous'' polarity and the local ST-FMR polarity is a strong indicator that the nonlocal signal contains a rectification component.

\paragraph*{(E) General trends.}
Table~\ref{tab:nonlocal_summary} condenses the role of the magnetic system parameters: a high dissipation favors inductive ST-FMR contributions in nonlocal readout, whereas a low dissipation gives rise to spin pumping even in nominal (local) ST-FMR geometries. Moreover, thicker films can reduce the influence of ST-FMR on both FMR and PSSW modes. Whereas many of the control experiments presented above are tests for detecting a potential contribution from inductive ST-FMR, the distance dependence from (B) seems the most robust procedure to prove a pure spin-pumping contribution, ruling out ST-FMR to unambiguously assign a magnon chirality or a spin Hall sign from a voltage polarity. Comparison against a local ST-FMR device (as in Fig.~\ref{fig3}) and the sign of the SSE is useful to control for the polarity of inductive effects.

\section{Conclusions}
\label{sec:Conclusions}
In conclusion, we have systematically investigated the coexistence and competition between spin pumping and ST-FMR in hybrid heavy-metal/ferrimagnetic-insulator heterostructures. By combining nonlocal (commonly interpreted as spin pumping) and local (commonly interpreted as ST-FMR) measurements on a series of Pt-capped garnet thin films with varied magnetic damping, we establish that the voltages detected in both types of experiments generally arises from a combination of each mechanism, with intrinsically opposite signs. In contrast to fully metallic Pt/ferromagnet bilayers, where spin-pumping contributions are only of the order of a correction to the actual ST-FMR signals~\cite{liu2021influence,aoki2022anomalous}, we find that in many occasions in Pt/ferrimagnetic-insulator bilayers, a sign change occurs. This anomalous sign is observed both due to spin pumping in local geometries intended for ST-FMR, or due to ST-FMR in non-local geometries intended for spin pumping.

The relative weights of those two contributions is mainly governed by spin-wave propagation length, device geometry, and magnetic configuration. We show that the spin pumping mediated by propagating spin waves dominates at short antenna–detector separations and in low-damping materials. In contrast, the ST-FMR contribution induced by antenna--detector inductive coupling dominates at larger distances or when the magnetization is tilted such that propagating-spin-wave excitation and detection are impeded.  By further reducing magnetic dissipation and accessing PSSW modes, we reveal that different spin-wave eigenmodes can selectively enhance either spin pumping or ST-FMR, depending on their spatial profiles. The distinct spatial decay of those two competing mechanisms, exponential for propagating spin waves and power-law for inductive coupling, provide a clear experimental means to disentangle them.

Our results clarify a long-standing ambiguity in the interpretation of electrical measurements of magnetization dynamics in magnetic insulators. They demonstrate that the microwave-induced voltage sign alone cannot be taken as a unique fingerprint of magnon chirality or spin Hall sign without carefully accounting for competing inner mechanisms. This work establishes a unified framework for understanding electrical detection in magnonic systems and provides practical guidelines for designing experiments and devices that reliably access specific spin-dynamics phenomena. We expect these insights to be broadly relevant for future magnonic and spin-orbitronic technologies based on low-damping magnetic insulators.

\begin{acknowledgments}
This research was supported by the Swiss National Science Foundation (Grant No.~200021-236524). 
H.W.~acknowledges the support of the China Scholarship Council (CSC, Grant No.~202206020091). 
H.M. acknowledges support from JSPS Postdoctoral Fellowship (Grant No.~23KJ1159) and Swiss Government Excellence Scholarships 2024-2025.
R.S.~acknowledges funding by the Deutsche Forschungsgemeinschaft (DFG, Grant No.~425217212). 
A.E.K acknowledges support from the ETH Zurich Postdoctoral Fellowship Programme (24-1 FEL-038). 
We acknowledge CNRS Physics for supporting our work through an international research project (IRP 2025 CORMORANT).

\end{acknowledgments}

\section*{Data Availability}

The data that support the findings of this article will be made openly available~\cite{DATASET}.

\appendix

\section{Micromagnetic Simulations}
\label{app:simulations}
To determine the sign of the spin-pumping voltage expected from propagating spin waves, micromagnetic simulations were performed using MuMax$^3$~\cite{vansteenkiste2014design}. The TmIG magnetic film with in-plane periodic boundary conditions had a size of $500 \times 500 \times 5~\mathrm{nm}^3$, discretized into cells of $5 \times 5 \times 5~\mathrm{nm}^3$, with the following material parameters: saturation magnetization $M_{\rm{s}} = 1.3 \times 10^5~\mathrm{A/m}$, exchange stiffness $A = 2.5~\mathrm{pJ/m}$, and damping constant $\alpha = 0.01$.

To model the regions with and without Pt capping, two separate simulations are performed with perpendicular anisotropy constants $K_{\rm{u}} = 1.76 \times 10^4~\mathrm{J/m^3}$ and $2.04 \times 10^4~\mathrm{J/m^3}$, respectively. The magnon spectra are obtained by performing a fast Fourier transform (FFT) of $m_x$ over $20~\mathrm{ns}$ after exciting the system with a sinc pulse field
\begin{equation}
\mathbf{h}_{\mathrm{sinc}} =
h_{\mathrm{sinc}}
\,\mathrm{sinc}(2\pi f_{\mathrm{sinc}} t)\,\hat{x},
\end{equation}
applied uniformly to the equilibrium states under different external magnetic fields.

By performing an inverse FFT at the resonance frequencies, the time-dependent magnetization dynamics is reconstructed, and the spin-pumping signal is evaluated as $(\mathbf{m} \times \dot{\mathbf{m}})_y$ (where $\dot{\mathbf{m}}$ is the time derivative) averaged over one precession period.

\begin{figure}[!htbp]
\hspace{-0.5cm}\includegraphics[width=60mm]{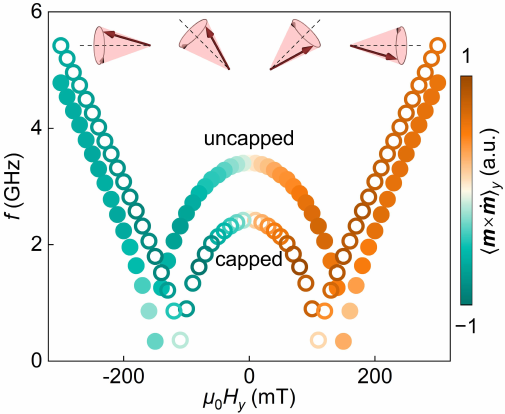}
\caption{\label{suppfig:fig5} Micromagnetic simulation results of field-dependent spin pumping signal of the capped (open symbols) and uncapped (filled symbols) regions. Insets show the magnetization precession with right-handed chirality at different external magnetic fields and orientations. }
\end{figure}

\section{Polar-Angle-Dependent Nonlocal Voltage  at Different  Frequencies}
\label{app:freqdep}
\begin{figure}[!htbp]
\hspace{-0.5cm}\includegraphics[width=90mm]{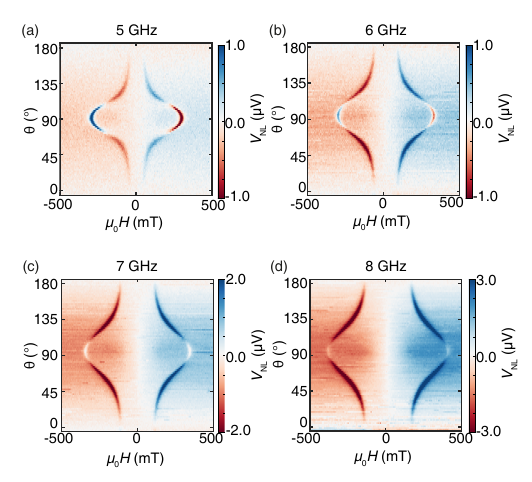}
\caption{\label{suppfig:fig2} Polar-angle dependent nonlocal voltage spectra under different excitation frequencies. For all panels, antenna--detector separation is 6~$\upmu$m, microwave power is +15~dBm.}
\end{figure}

In this section, we present polar-angle-dependent nonlocal voltage spectra measured at different excitation frequencies on a 5-nm-thick TmIG device with a separation distance of $s = 6~\upmu$m. As appears in Fig.~\ref{suppfig:fig2} (see also Fig.~\ref{suppfig:fig1}), the resonance linewidth increases with frequency, indicating enhanced magnetic dissipation and a reduced decay length for propagating spin waves. Correspondingly, the polar-angle range over which the ST-FMR contribution dominates becomes progressively broader with increasing frequency. These observations further confirm the competition between spin pumping and ST-FMR in the nonlocal voltage signal.

\section{PSSW Modes in Nonlocal Spin Pumping Voltages Spectra Measured on TmIG/YSGG Samples with Different Thicknesses}
\label{app:PSSW}
\begin{figure}[t]
\hspace{-0.5cm}\includegraphics[width=90mm]{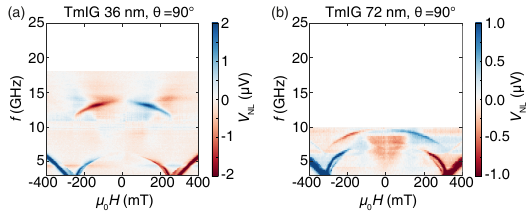}
\caption{\label{suppfig:fig4}IP field-dependent nonlocal voltage spectra measured on (a) 36-nm and (b) 72-nm-thick TmIG films grown on YSGG substrates, with an antenna--detector separation distance of 2~$\upmu$m. The power is fixed at +15~dBm. }
\end{figure}

In this section, we show that the nonlocal spin-pumping spectra obtained from TmIG/YSGG samples with three different thicknesses of 36 and 72~nm exhibit a clear opposite sign between the PSSW and FMR modes. The results are summarized in Fig.~\ref{suppfig:fig4}, where a systematic decrease of the upper mode frequency with increasing film thickness is observed, consistent with the expected thickness dependence of PSSW modes. This behavior indicates that for the electrical detection of PSSW modes, the ST-FMR mechanism typically plays a dominant role.

\bibliography{ref}

\end{document}